\documentclass{acadarticle}
\usepackage{graphicx}
\usepackage[hyphens]{url}
\newcommand{\jurnal}{Romanian Journal of Physics}
\newcommand{\volume}{\textbf{58}}
\newcommand{\numarul}{3-4}
\newcommand{\volyear}{2013}
\usepackage{mynatbib}
\title{The interplanetary mass ejections behaviour in the heliosphere}
\author{Cristiana Dumitrache}
\author{Nedelia A. Popescu}
\affil{\em Astronomical Institute of Romanian Academy\\%
\em Str. Cutitul de Argint 5,\\%
\em 040557 Bucharest, Romania\\ %
\em Email: crisd@aira.astro.ro; nedelia@aira.astro.ro}


\begin{document}
\maketitle

\begin{abstract}
We present here an overview of an important solar
phenomenon with major implications for space weather and planetary
life. The coronal mass ejections (CMEs) come from the Sun and expand
in the heliosphere, becoming interplanetary coronal mass ejections
(ICMEs). They represent huge clouds of plasma and magnetic fields
that travel with velocities reaching even 2000 km/s and perturbing
the planetary and interplanetary field. The magnetic clouds (MC) are a special class
of ICMEs. We summarize some aspects as the ICMEs identification, propagation and track back to the Sun, where
the solar source could be found. Each event has its own peculiarity. Much more, the ICMEs moving in the ecliptic plane are different from that travelling out of the Sun-Earth plane. Their study gives us an idea about the three dimension manifest of the heliosphere.
We notice here few known catalogs of the ICMEs and magnetic clouds, useful for the general studies of the ICMEs.
We also summarize some results of the authors previous work.
\end{abstract}

\section{Introduction}
Coronal mass ejections (CMEs) are huge clouds of magnetized plasma that erupt from the solar corona into the interplanetary space. They propagate in the heliosphere with velocities ranging from several hundred to several thousand km/s. Their signatures in the interplanetary space are different and their effects for space weather represent a keystone.

The CMEs are very important for the space weather conditions but also are the most important agent of the Sun influences on the planetary magnetospheres. Understanding of the onset and the solar source of these phenomena, as well as their traveling in space represent a challenge since each interplanetary coronal mass ejection (ICME) has its own characteristics.

There are many solar or heliospheric space missions that focused on the observations of these events. We remind here SOHO satellite very important for CMEs registration and ACE, WIND and Ulysses, satellites specialized for the ICMEs registration. While ACE and WIND missions have an orbit close to the ecliptic and register ICMEs that hit or pass near the Earth, Ulysses provided information about the 3D heliosphere. Launched in October 1990, Ulysses was the first mission to explore the heliosphere from the solar equator to the poles. The results of data analysis reveal that the solar magnetic field is carried out in space in a very complicated manner than previously expected. The northern polar pass of Ulysses' second orbit, in 2001, coincided with the phase of the solar cycle immediately following the sunspot maximum and the other north solar pole overpass corresponded to the solar minimum activity, in 2007. Ulysses spacecraft ended its mission in 2009.

Ulysses in-situ measurements have given a new picture on the large-scale properties of the fast solar wind as well as on the microscopic scales. From the large sample of data obtained during Ulysses fast latitudinal scans in 1994 and 2007 occurring both near solar minima, \cite{Issautie2010} analyzed the spectrum of the electron density fluctuations. In a special review \cite{Issautie2008} synthesized the main Ulysses results during its entire mission.

It is generally known that the characteristics of the magnetic field at low and high latitudes are different, at least for the reason of solar activity features that manifest at that latitudes, active regions and polar filaments, respectively. Many CMEs originating at low latitude could arrive at the Earth’s orbit and at the ecliptic level. Those occurred at high latitude cannot be intercepted by the majority of spacecraft flying near Earth.

Beyond one astronomical unity, the ICME identification is more difficult because some ICME signatures are blurred through interaction with the ambient solar wind. For this reason the observations provided by Ulysses spacecraft during its transient over the solar poles are very important to help us understand how the high-latitude CMEs travel through space, far from the ecliptic and beyond 1 AU, and have an insight to the 3D heliosphere.

Many ICMEs display magnetic clouds (MCs) in their structure. A MC represents the interplanetary manifestation of a flux rope expelled from the Sun. Magnetic clouds were first defined by \cite{Klein1981} as regions with a radial dimension of approximately 0.25 AU at 1 AU, in which the magnetic field strength is high. Smoothly changing of the field direction is observed when a spacecraft passes through the cloud. Combined with the low proton temperature, the strong magnetic field leads to a low proton $\beta< 0.2$. In a recent study, \cite{Du2010} found that about $43\%$ of ICMEs were MCs and $23\%$ were associated with radial events.

Many authors studied specific events tracking a CME through the interplanetary space until they reached a certain spacecraft or even using more spacecraft. We point out here the works of \cite{Dasso2005}, \cite{Dasso2006}, \cite{Dasso2007}, \cite{Dasso2009} and \cite{Dassoetal2009}. We also notice important review papers on ICMEs: \cite{Demoulin2008b}, \cite{VDGesztelyi}, \cite{Schmieder2011}.

\section{The ICMEs identification}

An interplanetary mass ejection (ICME) could be identified in the solar wind plasma by a sum of characteristics regarding its magnetic field, velocity,
temperature and density. In a sum of articles, Dumitrache \etal ( \cite{Dumitrache2009},\cite{Dumi2010}, \cite{DumiPope2010a}, \cite{DumiPope2010b}, \cite{Dumitrache2010c}, \cite{Popescu2010AIPC1216} and \cite{Dumitrache2011SoPh}) used observations provided by three instruments on-board of Ulysses spacecraft for few ICME events identification. From SWOOPS they used hourly averages data for solar wind plasma bulk parameters (i.e. velocity (\emph{v}), proton density (\emph{Np}), proton temperature (\emph{Tp}), \emph{alpha/proton} abundance ratio,
$N(He++)/N(H+))$. From VHM magnetic field instrument, they used also hourly data for the magnetic field magnitude (\emph{B}), respectively for the magnetic
field components $(Br,Bt,Bn)$, given in units of nT (nanoteslas) and in RTN coordinates, where R is the sun-s/c axis, T is the cross product of the
solar rotation axis and R, and N is the cross product of R and T. From SWICS, the solar wind ion composition analyzer instrument, the authors used
3 hour averaged data for temperature, plasma composition and charge-state measurements of solar wind ions.

The interplanetary mass ejections are commonly identified by a set of signatures like  (\cite{Richardson1993}, \cite{Gazis2006}, \cite{Zurbuchen2006}, \cite{Ebert2009}):

\begin{description}
\item - alpha (He++) abundance enhancement;

\item - low ion temperatures during the event;

\item - heavy ion species present anomalies of abundance and charge state;

\item - bi-directional electron streaming moving in both directions along
the magnetic field lines;

\item - increased intensity of the magnetic field;

\item - forward and reverse shocks in the magnetic clouds case;

\item - smooth magnetic field rotation in the magnetic clouds case.
\end{description}

The helium (alpha particles) abundance is considered by \cite{Steiger2006} as being  the best signature suitable to track an ICME in the heliosphere. An
ICME occurrence is determined \emph{in situ} by helium abundance enhancement: the alpha particles density rated to hydrogen density should be greater than
$0.08$  (\cite{Neugebauer1997a}).

The low ion temperature represents a signature for magnetic clouds too (\cite{Richardson1993}). The ratio of the measured proton temperature to
that of expected proton temperature for the normal solar wind ($Tp/T\exp $ ) represents a quantifiable ICME signature. This ratio should be less than $0.5$,
and is multiplied by $10^{3}$K. The expected temperature was computed for distances exceeding 1 AU  (\cite{Wang2005}) by:

\[
T\exp =\left\{
\begin{array}{c}
\frac{(0.031v-5.1)^{2}}{r^{0.7}},v<500 \\
\frac{0.51v-142}{r^{0.7}},v\geq 500%
\end{array}%
\right.
\]

\begin{figure}[h!tb]
\begin{center}
\includegraphics[scale=0.8]{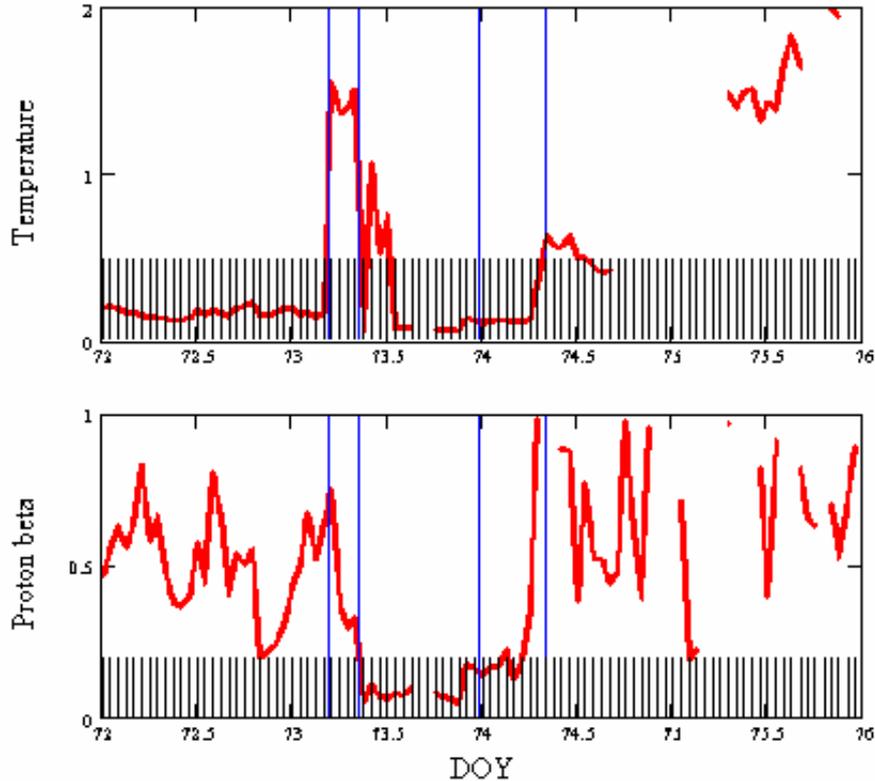}%
\end{center}
\caption{Temperature and proton beta decreasing under a threshold
during the ICME event on 14 March 2001. } \label{f1}
\end{figure}

During an ICME event the plasma $\beta $ significantly decreases, especially inside a MC. A protons beta plasma was computed using the formula: $\beta
=p/(B^{2}/2\mu )$. This parameter is less than 0.2 in magnetic clouds, thus revealing a force free magnetic field and the presence of solar coronal
matter inside the magnetic clouds. Fig. \ref{f1} displays these parameters during the Ulysses ICME event registered on 1 March 2001. The vertical long lines
represent the magnetic cloud borders, where the first region is the forward shock, the middle is the magnetic cloud itself and the last one is the
reverse shock region. The small vertical lines indicate the  threshold values admitted for an event to be an ICME.

Elemental composition represents an important indicative of solar wind source and for the ICMEs presence too. \cite{Burgi1986} stated that the
charge state information within an ICME reflects the temperature history of the CME solar counterpart. Charge state composition of elements such as O,
C, and Fe, contains a coronal signature of the solar wind and extremely high charge states are associated with mass ejections. Very high charge states of
Fe are considered the most reliable signature of an ICME (\cite{Lepri2001}). The $Fe/O$ abundance ratio divided by the solar $Fe/O$ ratio (about 0.05) represents a proxy for the strength of the first ionization potential fractionation. Weighted average of all charge states from $Fe(6+)$ to $Fe(16+)$ are normally between 9 and 11 in the solar wind and higher values indicate the presence of an ICME (\cite{Lepri2004}). Another signature of an ICME is the fulfilling of the condition $O(7+)/O(6+)>1$. In their study Henke \etal (\cite{Henke98} asserted that an enhanced $O(7+)/O(6+)$ density ratio is directly correlated with magnetic clouds.

According to \cite{Burlaga1977JGR....82.3191B}, the magnetic clouds are also characterized by a smooth coherent rotation of the magnetic field parallel to a plane, when the
spacecraft goes through the ICME considered as a flux rope. This rotation can occur in any direction on a time scale from few hours to days. These
changes can be detected in the spacecraft Cartesian components of the magnetic field ($Bx,By,Bz$), where $Bx$ points from the spacecraft towards
the Sun, $By$ points towards the East, and $Bz$ points normal to them towards North. The directional changes of the magnetic field can be
investigated with the minimum variance method (MVA) applied to the magnetic field components in these coordinates.  \cite{Klein1981} or \cite{Bothmer1998} described in detail the MVA method applied to MCs.

As the CMEs propagate in the heliosphere, their internal properties and configuration change and extend, so it is difficult to relate back an ICME registered at several AU. One parameter that can be measured in-situ by satellites remains yet unchanged: this is the ionization level of the solar wind. By measuring charge states of solar wind ions, thermodynamic properties present in the source region of the solar wind can be analyzed at any distance in the heliosphere. The charge state is higher inside most of
ICMEs, especially inside the magnetic clouds. \cite{Rodriguez2004} extended the analysis  made by \cite{Henke98} to a set of 40 magnetic clouds detected by Ulysses and found that the increase of the charge state is present at all latitudes and phases of solar cycle. \cite{Wurz2001} detected the elemental composition in magnetic clouds observed in 1997 and 1998. They observed that the heavy elements (carbon through iron), which can be regarded as tracers in the solar wind plasma, display a mass-dependent enrichment
of ions monotonically increasing with mass. When comparing the MC plasma to regular solar wind composition, a net depletion of the lighter ions, helium through oxygen, was always observed.

\section{The ICMEs propagation}

One way to better understand how ICMEs evolve in the solar wind is to track specific ICMEs observed by spacecraft at different heliospheric distances.
\cite{Wang2005} described the most important characteristics of the interplanetary coronal mass ejections in the heliosphere between 0.3 and 5.4 AU, using data from several spacecraft.

An excellent work was also done by \cite{Foullon2007} that investigated
in detail a low latitude event coming from an active region filament and
registered by more than one spacecraft. In the high-latitude ICME case we
cannot perform a follow up using satellites like Ace, Wind or Helios 1
since the event is not near the ecliptic. Using data from Ulysses
we could track an event occurred at high latitude and far from Earth orbit
(1AU), at times when the satellite passed over the solar poles. The latitude
distribution of the CMEs occurrence is solar cycle dependent and this
implies that only during some periods we could expect to observe high
latitude CMEs and ICMEs.

Many authors derived empirical models from observational data, concerning
the CMEs propagation through heliosphere. \cite{Lindsay1999} and
\cite{Gopalswamy2001} developed simple shock time of arrival models using
the initial CME speed and ambient solar wind speed to predict the
acceleration or deceleration of the CME. \cite{Gopalswamy2001} assumed
that the CME speed remains constant from the Sun to 1 AU.
\cite{Owens2004} analyzed the existing models of ICMEs travel between Sun and Earth
and found a similar accuracy of the 1 AU arrival predictions of all models,
eventual errors being due to the CME geometry of trajectory and expansion.

Another way to track an ICME is to observe the associated CME with a
coronagraph at the solar limb and the in situ ICME with a spacecraft in
quadrature with the coronagraph. It means to follow up a CME in space until
it reaches a spacecraft and we get recorders from this. \cite{Demoulin2009}
showed that the slow CMEs (with $v<400$ km/s) accelerate and reach the solar wind
speed, while the most rapid CMEs slow down and tend, to long distance, to
the solar wind ambient speed. A similar idea is accounted by \cite{Borgazzi2008} who also  took into consideration the drag force between the ICME and
the solar wind. Typically the strongest deceleration occurs close to the
Sun, and ICMEs have a nearly constant velocity in most of the heliosphere.
\cite{Demoulin2008a} analyzed the expected \emph{in situ} velocities from a
hierarchical model for expanding interplanetary coronal mass ejections and
showed that the global acceleration of ICMEs has, at most, a small
contribution to the in situ measurements of the velocity. He also
analytically proved that MCs are expanding at a comparable rate,
independently of their size or field strength, despite very different
magnitudes in their velocity profiles.

\begin{figure}[h!tb]
\begin{center}
\includegraphics[scale=0.5]{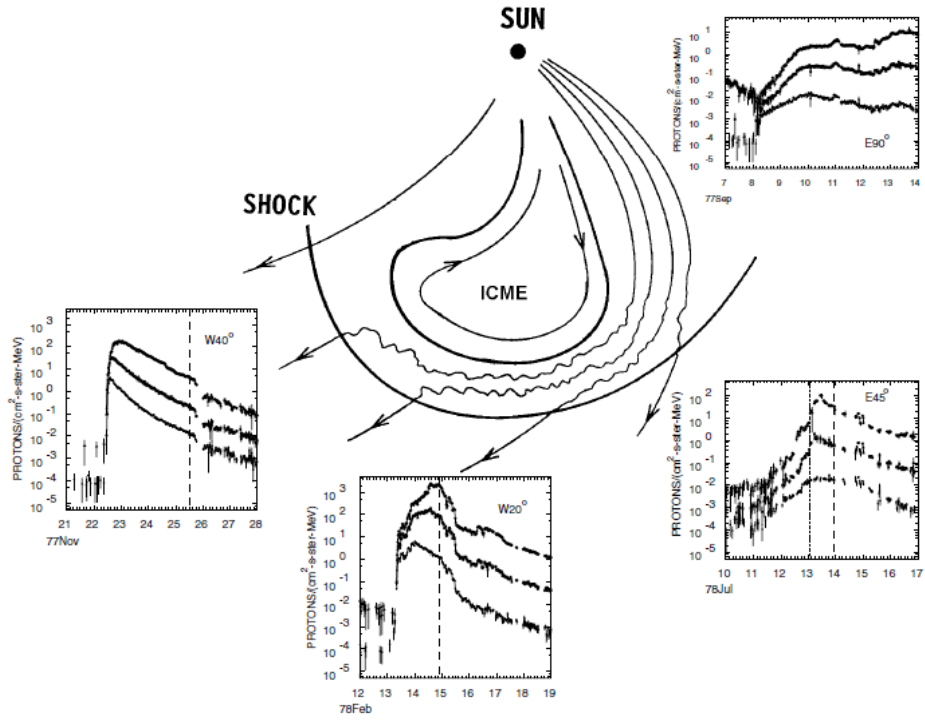}
\end{center}
\caption{The shape of an ICME and surrounding interplanetary field structure including
the presence of a shock (Cane et.al., 1988).}
\label{cane88}
\end{figure}

Observations have proved that a coronal mass ejection expand into heliosphere
as it travel. Simultaneous observations by widely separated spacecraft show
that in large events particles reach widespread regions of the heliosphere,
up to $300^{o}$ in longitude (\cite{Cliver1995}); and up to at least $80^{o}$ in latitude (\cite{Lario2003}). Energetic particle observations
from interplanetary spacecraft can be used to infer the properties of the traveling
CMEs. \cite{Cane1988} have shown that the proton intensity profiles of the
solar energetic particle (SEP) events observed in the ecliptic plane at 1 AU are organized in terms of
the longitude of the observer with respect to the traveling CME-driven shock,
and resume their results in Fig. \ref{cane88}.

However at large heliocentric distances and at high heliolatitudes the
relation between the origin of the event and the time-intensity profiles is
less clear, so this result can not be applied to Ulysses observations.

Observations of the energetic particles during the passage of an ICME over
the observer (spacecraft) provide valuable information about the structure
of the ICME. We should distinguish the signatures of an ICME at the ecliptic
level and one astronomical unity (near Earth) and those at high latitude and
long distances in heliosphere. For example, an ICME\ passage in the ecliptic
plane is associated with the energetic particle intensity depletion, known
as the Forbush decreases (\cite{Cane2000}). In contrast to this, the
high-latitude ICMEs display an enhancement of the particle intensities.

\section{The tracking back to the Sun}

The track back to the Sun of an ICME registered at distances greater than 1
AU still remains a challenge. Several authors have identified the solar
counterparts of Ulysses' ICMEs using different methods. We remind here a
few: energetic particle flux (\cite{Watari2002cosp...34E2490}; \cite{Simnett2003ESASP.535..613S}),
flux rope modeling (\cite{Demoulin2008a}), tracking the
interplanetary scintillation close to the Sun and out of ecliptic (\cite{Tokumaru2006}). \cite{Gazis2006} provided a list of works about Ulysses'
ICMEs and their solar counterparts' determination, but this list contains
events occurred until 2001.

\begin{figure}[h!tb]
\begin{center}
\includegraphics[scale=0.8]{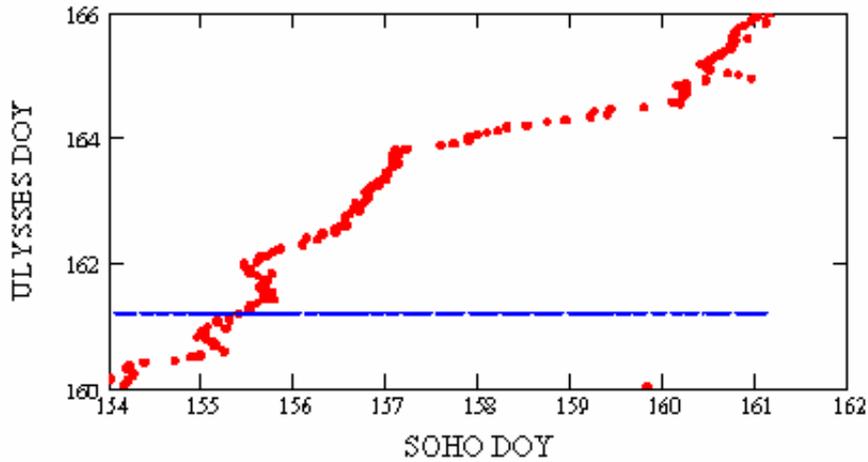}
\end{center}
\caption{The track back to the Sun diagram of the ICME event from 10
June 2001 registered by Ulysses spacecraft. (\cite{DumiPope2010b}).  } \label{f3}
\end{figure}

\cite{Dumi2010}, \cite{DumiPope2010a}, \cite{DumiPope2010b}, \cite{Dumitrache2010c}, \cite{Dumitrache2011SoPh} treated the problem of the track
back to the Sun of some ICME events registered by Ulysses spacecraft and
identified their solar sources. They considered a linear kinematic law to
track the ICME back to the Sun using the velocity registered in situ by
Ulysses. Therefore they introduced a graphical method to detect the day of
the year when a specific event occurred on the Sun. Fig. \ref{f3} plots a
diagram with the days of the year of events registered by Ulysses on
vertical axis, while on the horizontal axis are the days of the year
of events should appear on the Sun and registered by SOHO. Using the CDAW
catalog the authors investigated more CMEs candidates as being the solar
counterpart of a specific ICME event. The dashed horizontal line marks the
DOY of the ICME registered by Ulysses, so the intersection of this one with
the curve of the events must give on abscise the day of the solar
counterpart apparition. They also performed the follow up of all CMEs
candidates registered by SOHO spacecraft until their Ulysses' arrival. This
follow up was based on a linear model but using two different velocities:
first, the initial linear CMEs' velocity, and, second, a velocity empirical
formula fitted by \cite{Lindsay1999}, that can be considered as the CMEs
are embedded in the solar wind. Both computations indicated the same solar
counterpart and the errors are of the order of hours. Formula deduced by
\cite{Lindsay1999}, i.e. the radial velocity ($vr$) is deduced from the initial
coronograph measured velocity ($vi$), $vr=vi\cdot 0.25+360$, gives more accurate
results.

\cite{Lindsay1999} studied statistical relationships between the speeds
of coronal mass ejections observed near the Sun and key characteristics of
the associated interplanetary disturbances (ICMEs) detected near the
ecliptic at distances less than 1 AU, using a combination of Solwind and
SMM coronagraph data and Helios-1 and Pioneer Venus Orbiter interplanetary
field and plasma. These observations confirm that while the occurrence of
large interplanetary magnetic field magnitudes and high bulk plasma speeds
associated with ICME passage may be predictable from coronagraph-derived CME
speeds, other important ICME features like large-magnitude southward $Bz$
require other diagnostics and tools for forecasts.

\section{About the solar sources}

A major step to understand the variability of the space environment is to
link the sources of coronal mass ejections (CMEs) to their interplanetary
counterparts or vice versa, to track back to the Sun the interplanetary
disturbances and found their solar source.

A coronal mass ejection has different signatures in its core region of
apparition as well as large-scale signatures by the neighborhood structures
involved in. A CME can be detected by the occurrence of filament eruption or
flare loop arcade, and X-ray/EUV double dimming. These signatures are
observable in many wavelengths, but especially in H$\alpha$, EUV, soft X-ray
and/or radio. Disturbances along coronal hole boundaries have also been
reported as possible signatures of CMEs (\cite{Hudson1996}; \cite{Attrill2006}; \cite{Veronig2006}; \cite{Harra2007}).

The principal sources of CMEs are the filaments or prominences. Very often
the prominences form in more complex structure where these ones are situated
at the base. Above it, a coronal streamer build up, separated by a cavity.
Considering the filaments classifications, we could distinguish more type of
sources for the mass ejections, i.e. CME starting from polar filaments, from
active regions filaments or from complex filaments.

Flares represent another important source of CMEs. Here we could distinguish
also few aspects: there are flares occurring in active regions and spotless
flares. However, the basic process of CMEs sources is linked to magnetic
field change from quasi-static state to an erupting stage.

We remind here an interesting statistics made by \cite{Wenbin2007}:
they analyzed the solar cycle variation of the real CME latitudes. Their
conclusions are that high-latitude CMEs constituted only $3\%$ of all CMEs
and mainly occurred during the period of solar polar magnetic field
reversal, while $4\%$ of all CMEs occurred in between $45^0$ and $60^0$ latitude.
These mid-latitude CMEs occurrence presented three peaks in 2000, 2002 and
in 2005. The highest occurrence rate of low-latitude CMEs (with latitude
less than $45^0$) was at the maximum and during the declining phase of the
solar cycle 23. The latitudinal evolution of low-latitude CMEs does
not obey the Sp\"{o}rer sunspot law and this fact suggests that many CMEs
originated outside of active regions and consequently to be eruptive
prominences or spotless flares signatures.

\cite{Vilmer2003} analyzed and identified their solar sources for 32
interplanetary disturbances that affected the Earth. They found a large
proportion to be associated with flares originating in active regions and
only ten cases were associated to filament activity situated at the disk
center. Other statistics, considering generally the CMEs, found that
filaments are mostly sources for these explosive phenomena. \cite{VDGesztelyi} synthesized the magnetic clouds
characteristics and their relation to the solar sources in a comprehensive
work.

\cite{Gopalswamy2010} identified a large number of interplanetary
disturbances that do not have a discernable driver by the spacecraft near
Earth observations. These shocks are associated with fast and wide CMEs and
radio bursts type II. These CMEs were located near coronal holes. The
CME-coronal hole interaction must be widespread in the declining phase of
the cycle 23 and may have a significant impact on the geoeffectiveness of
CMEs.

\cite{Schmieder2006JApA...27..139} analyzed the magnetic source regions of coronal mass
ejections and revealed mainly two of them: the decaying active regions and
the filaments. Usually an ICME keeps its solar source magnetic helicity.
However, \cite{Chandra2010} reported the observation of a solar source
with negative helicity generating a magnetic cloud with positive polarity.
This magnetic cloud gave the largest geomagnetic storm of the solar cycle 23.

\section{Conclusions}

Many authors studied specific CME-ICMEs events and found many general
properties of their in-situ identification, their propagation and evolution,
but there are also many others things that make unique each event. There are
also many unsolved aspects of the ICMEs diagnostic. A special class of the
ICMEs is represented by the magnetic clouds, which are very important for
their space weather amplified effects. The ICMEs were registered in situ by
many satellites and there are few catalogues that attempted to link them to
their solar sources, but the majority of these satellites are near the
ecliptic and one astronomical unity (near Earth). Only Ulysses spacecraft
had a unique orbit allowing registrations far from the ecliptic plane and
beyond 1 AU, giving us a 3D view of the heliosphere. The tracking of a CME
far than 1 AU is not an easy task and is more complicated when the events
are at high-latitudes and no one additional registrations between the Sun
and spacecraft can be found. For this reason the results of our PECS-ESA
project (2007-2009) that focused on all these aspects represented an
important step in the understanding of the CME-ICMEs events. We have
analyzed in a series of articles more ICME events and their solar sources:
10 June 2001, 18 August 2001, 24 August 2001, 18 January 2002, 14 March
2002, 5 May 2002, and 15 July 2007.


\end{document}